# Efficient Independent Vector Extraction of Dominant Target Speech

Lele Liao, Zhaoyi Gu, Jing Lu

*Abstract*—The complete decomposition performed by blind source separation is computationally demanding and superfluous when only the speech of one specific target speaker is desired. In this paper, we propose a computationally efficient blind speech extraction method based on a proper modification of the commonly utilized independent vector analysis algorithm, under the mild assumption that the average power of signal of interest outweighs interfering speech sources. Considering that the minimum distortion principle cannot be implemented since the full demixing matrix is not available, we also design a one-unit scaling operation to solve the scaling ambiguity. Simulations validate the efficacy of the proposed method in extracting the dominant speech.

## I. Introduction

BLIND speech separation (BSS) [1] is a fundamental task in speech processing, aiming at separating target speech from background interference. For single-microphone (mono-channel) processing, the data-driven deep learning based solution [2]-[4] has attracted many researchers' attention and been proven to be a promising technique. However, the generalization of the supervised end-to-end deep learning approach to untrained condition is still a crucial issue, especially when speech signals of more than 2 speakers coexist in the mixture. For multi-microphone (multi-channel) processing, the independent vector analysis (IVA) with its variation [5]-[9] can be regarded as the state-of-the-art solution without any training processes, and is of particular interest in this paper. IVA has recently been integrated with the multichannel nonnegative matrix factorization (MNMF) [10], [11] scheme, resulting in a unified method called the independent low-rank matrix analysis (ILRMA) [12], [13].

In many practical applications, only speech of one specific speaker is the source of interest (SOI). For commonly utilized BSS, it is necessary to separate all sources and find the desired one afterwards. It is obvious that the calculation for the undesired signals is in essence redundant, not to mention the challenging task of finding the desired speech from the multiple separated signals. Blind speech extraction (BSE) [1], [14]-[16] is an efficient alternative to BSS, aiming at extracting the desired signal from the mixture directly. Cichocki *et al.* proposed in [17] the method that extracts the source signal with maximal absolute normalized kurtosis value, which can be extended to extract the source signal whose kurtosis value lies in a specific range [18]. The approach in [19] is based on maximal correlation with an explicit template of SOI. Moreover, the sparseness [20], the temporal structure [21], [22] and the frequency structure [23] can also be utilized to extract SOI in BSE. These algorithms all impose strong assumptions on SOI, therefore their practical applications are limited. The recently developed methods in [24], [25] do not have strict assumption on the SOI model, but require the prior knowledge on the mixing or demixing parameters, which can hardly be obtained in many practical applications. Another intriguing BSE method is proposed in [26], which can be extended to multi-source extraction scenario [27]. The Gaussian distribution assumption of the interference also challenges its performance in practical applications.

In this paper, we design an efficient BSE method by modifying IVA using the negentropy-based cost function with several different source prior models. Only the demixing vector related to SOI is updated with fast fixed-point method, resulting in an algorithm called fast independent vector extraction (FastIVE). The algorithm is initialized by principal component analysis (PCA), with mild assumption that the SOI is from the dominant speaker in the mixture, whose speech power at the microphones outweighs the interferences. Since the demixing matrix is not available during the iterative process, we also design a scaling operation using only the SOI demixing vector to solve the scaling permutation problem.

## II. The Proposed Algorithm

### A. Problem Formulation

The mixed signal of $N$ sources captured by $M$ sensors after short time Fourier transform (STFT) can be expressed as

$$\mathbf{x}^k = \mathbf{H}^k \mathbf{s}^k, \qquad (1)$$

where the superscript $k$ denotes the index of frequency, $\mathbf{s}^k = [s_1^k, \ldots, s_N^k]^\mathrm{T}$ is the source vector, $\mathbf{x}^k = [x_1^k, \ldots, x_M^k]^\mathrm{T}$ is the

This work was supported by the National Natural Science Foundation No. 11874219 of China.

Lele Liao Zhaoyi Gu and Jing Lu are with Key Laboratory of Modern Acoustics, Institute of Acoustics, Nanjing University, Nanjing 210093, China (e-mail: liaolele@smail.nju.edu.cn; guzhaoyi@smail.nju.edu.cn; lujing@nju.edu.cn).

mixture vector, the superscript T denotes the matrix transpose operation, and $\mathbf{H}^k$ is the mixing matrix.

IVA's cost function is based on mutual information [28]

$$I\{\mathbf{y}_1...\mathbf{y}_N\} = \sum_{n=1}^{N} H\{\mathbf{y}_n\} - \sum_{k=1}^{K} \log|\det \mathbf{W}^k| - H\{\mathbf{x}^1...\mathbf{x}^K\}, \quad (2)$$

where $\mathbf{W}^k$ is the demixing matrix whose $n$-th row is denoted by $\left(\mathbf{w}_n^k\right)^H$, $\mathbf{y}_n \left(= \left[y_n^1, y_n^2, ..., y_n^K\right]^T\right)$ is the output signal with $y_n^k = \left(\mathbf{w}_n^k\right)^H \mathbf{x}^k$ (the superscript H denotes the complex conjugate transpose operation), and the notation of $H\{\cdot\}$ stands for the entropy. Note that $H\{\mathbf{x}^1...\mathbf{x}^K\}$ equals to a constant, so the cost function can be simplified as

$$C_{IVA} = \sum_{n=1}^{N} H\{\mathbf{y}_n\} - \sum_{k=1}^{K} \log|\det \mathbf{W}^k|. \quad (3)$$

The determinant of $\mathbf{W}^k$ is a coupling term of all demixing filters that makes it mandatory to update the whole matrix $\mathbf{W}^k$, resulting in redundant calculation when only one SOI needs to be extracted.

*B. BSE with IVE*

Instead of mutual information, we design the cost function of IVE similar to IVA but in the form of negentropy as

$$C_{IVE} = \sum_{n=1}^{N} N\{\mathbf{y}_n\} = \sum_{n=1}^{N} \left(H\{\mathbf{y}_n^{Gauss}\} - H\{\mathbf{y}_n\}\right), \quad (4)$$

where $\mathbf{y}_n^{Gauss}$ is a Gaussian variable with the same mean and variance as $\mathbf{y}_n$, and the term $H\{\mathbf{y}_n^{Gauss}\}$ is actually constant and negligible, so that (4) can be further simplified as

$$C_{IVE} = -\sum_{n=1}^{N} H\{\mathbf{y}_n\} = \sum_{n=1}^{N} \tilde{E}\left\{\log\left[f_{\mathbf{s}_n}(\mathbf{y}_n)\right]\right\}, \quad (5)$$

where $f$ denotes the probability distribution function (PDF) of the source signals and $\tilde{E}[\cdot]$ is the sample-based estimate of expectation. Setting $G = -\log\left(f_{\mathbf{s}_n}(\mathbf{y}_n)\right)$ as a function of $\|\mathbf{y}_n\|^2$ yields

$$C_{IVE} = -\sum_{n=1}^{N} \left[\tilde{E}\left[G\left(\sum_{k} \left|\left(\mathbf{w}_n^k\right)^H \mathbf{x}^k\right|^2\right)\right]\right]. \quad (6)$$

Note that the cost function of (6) in theory is the summation of aggregated statistics concerning nongaussianity of $N$ independent sources, which makes it possible to search for the maximum for a single projection $\mathbf{y}_n$ and extract only one desired source. The problem remains is that which source should be extracted. In practical applications, the SOI often has a higher signal power than the interference (e.g., the SOI is the closest to the array among all the speakers, or the desired speaker intentionally increases volume in noisy conditions), and in this case it is reasonable to apply PCA to firstly identify the component with maximal variance (power) and then exert the IVE.

Following the standard procedure of PCA[29], the input vector $\mathbf{x}^k$ is transformed as

$$\tilde{\mathbf{x}}^k = \left(\mathbf{U}^k\right)^T \mathbf{x}^k. \quad (7)$$

where $\mathbf{U}^k$ consists of the eigenvectors corresponding to the eigenvalues of the covariance matrix $\mathbf{C}_x$ of $\mathbf{x}^k$ in descending order. Obviously the first component of $\tilde{\mathbf{x}}^k$ has the maximal signal power. For the purpose of extracting the dominant component, the cost function can be simplified as

$$C_{IVE}\left(\mathbf{w}_1^k\right) = -\tilde{E}\left[G\left(\sum_{k} \left|\left(\mathbf{w}_1^k\right)^H \tilde{\mathbf{x}}^k\right|^2\right)\right]. \quad (8)$$

To accelerate convergence, we further whiten the input data and transform (7) into

$$\tilde{\mathbf{x}}^k = \left(\mathbf{D}^k\right)^{-1/2} \left(\mathbf{U}^k\right)^T \mathbf{x}^k, \quad (9)$$

where $\mathbf{D}^k = \text{diag}(d_1^k, ..., d_N^k)$ is a diagonal matrix whose diagonal elements are eigenvalues of $\mathbf{C}_x$ in descending order. The demixing vector $\mathbf{w}_1^k$ is initialized as an one-hot vector $\mathbf{e}_1$ whose first element is 1 so that the component with the maximal power is obtained. Under the assumption that the SOI has the highest signal power, $\tilde{x}_1^k$ can be regarded as the component dominated by the SOI, and $\mathbf{w}_1^k$ initialized as $\mathbf{e}_1$ is spatially closer to the optimal point w.r.t the SOI rather than other interference sources and easily converges to the demixing filter of the SOI.

Similar to the fast fixed-point algorithm derived in [6], applying Taylor series polynomial approximation to the derivative of the cost function shown in (8) yields

$$\frac{\partial C_{IVE}\left(\mathbf{w}_1^k\right)}{\partial\left(\mathbf{w}_1^k\right)^*}$$
$$\approx \frac{\partial C_{IVE}\left(\mathbf{w}_{1,o}^k\right)}{\partial\left(\mathbf{w}_1^k\right)^*} + \frac{\partial^2 C_{IVE}\left(\mathbf{w}_{1,o}^k\right)}{\partial\left(\mathbf{w}_1^k\right)^* \partial\left(\mathbf{w}_1^k\right)^T}\left(\mathbf{w}_1^k - \mathbf{w}_{1,o}^k\right) \quad (10)$$
$$+ \frac{\partial^2 C_{IVE}\left(\mathbf{w}_{1,o}^k\right)}{\partial\left(\mathbf{w}_1^k\right)^* \partial\left(\mathbf{w}_1^k\right)^H}\left(\mathbf{w}_1^k - \mathbf{w}_{1,o}^k\right)^* \equiv \mathbf{0}$$

Note that the subscript 'o' indicates the parameters of the current iterate. Let $G'(\cdot)$ and $G''(\cdot)$ denote the first and second order derivatives of $G(\cdot)$ respectively, then the terms of the right hand side of (10) become

$$\frac{\partial C_{IVE}\left(\mathbf{w}_{1,o}^k\right)}{\partial\left(\mathbf{w}_1^k\right)^*} = \tilde{E}\left[\left(y_{1,o}^k\right)^* G'\left(\sum_{k}\left|y_{1,o}^k\right|^2\right)\tilde{\mathbf{x}}^k\right], \quad (11)$$

$$\frac{\partial^2 C_{IVE}\left(\mathbf{w}_{1,o}^k\right)}{\partial\left(\mathbf{w}_1^k\right)^* \partial\left(\mathbf{w}_1^k\right)^T}$$
$$= \tilde{E}\left[\left(G'\left(\sum_{k}\left|y_{1,o}^k\right|^2\right) + \left|y_{1,o}^k\right|^2 G''\left(\sum_{k}\left|y_{1,o}^k\right|^2\right)\right)\tilde{\mathbf{x}}^k\left(\tilde{\mathbf{x}}^k\right)^H\right] \quad (12)$$
$$\approx \tilde{E}\left[G'\left(\sum_{k}\left|y_{1,o}^k\right|^2\right) + \left|y_{1,o}^k\right|^2 G''\left(\sum_{k}\left|y_{1,o}^k\right|^2\right)\right]\tilde{E}\left[\tilde{\mathbf{x}}^k\left(\tilde{\mathbf{x}}^k\right)^H\right]$$
$$= \tilde{E}\left[G'\left(\sum_{k}\left|y_{1,o}^k\right|^2\right) + \left|y_{1,o}^k\right|^2 G''\left(\sum_{k}\left|y_{1,o}^k\right|^2\right)\right]$$

and

$$\frac{\partial^2 C_{IVE}(\mathbf{w}_{1,o}^k)}{\partial(\mathbf{w}_1^k)^* \partial(\mathbf{w}_1^k)^H}$$

$$= \tilde{E}\left[\left(\left(y_{1,o}^k\right)^*\right)^2 G''\left(\sum_k |y_{1,o}^k|^2\right) \tilde{\mathbf{x}}^k (\tilde{\mathbf{x}}^k)^T\right] \quad (13)$$

$$\approx \tilde{E}\left[\left(\left(y_{1,o}^k\right)^*\right)^2 G''\left(\sum_k |y_{1,o}^k|^2\right)\right] \tilde{E}\left[\tilde{\mathbf{x}}^k (\tilde{\mathbf{x}}^k)^T\right]$$

$$= \mathbf{0}$$

because of the commonly used circularity assumption $\tilde{E}\left[\tilde{\mathbf{x}}^k (\tilde{\mathbf{x}}^k)^T\right] = \mathbf{0}$.

Plugging (11), (12) and (13) into (10), we obtain the updating rule of FastIVE as

$$\mathbf{w}_1^k \leftarrow \tilde{E}\left[G'\left(\sum_k |y_1^k|^2\right) + |y_1^k|^2 G''\left(\sum_k |y_1^k|^2\right)\right] \mathbf{w}_{1,o}^k \\ - \tilde{E}\left[\left(y_1^k\right)^* G'\left(\sum_k |y_1^k|^2\right) \tilde{\mathbf{x}}^k\right]. \quad (14)$$

The notation "←" means updating the left subject by the right-hand terms. After each iteration, $\mathbf{w}_1^k$ is normalized by

$$\mathbf{w}_1^k \leftarrow \frac{\mathbf{w}_1^k}{\|\mathbf{w}_1^k\|}. \quad (15)$$

For the commonly used spherically symmetric Laplacian (SSL) [30] source prior model, $G(z) = \sqrt{z}$, $G'(z) = 1/2\sqrt{z}$, $G''(z) = -1/4\sqrt{(z)^3}$. The block permutation problem of IVA has been noted [8], and several improved models can be introduced into IVE. The multivariate generalized Gaussian source prior (GG) [30] has heavy tails, making it more suitable for modeling nonstationary signals like speech. For this model, $G(z) = z^{1/14}$, $G'(z) = 1/(14 z^{13/14})$ and $G''(z) = -13/(196 z^{27/14})$. Another proper choice is the multivariate student's t-distribution [8] that models the dependence structure more accurately with score function expressed as $G(z) = \log(1 + z/v)$, where $v$ is the degree of freedom parameter. Correspondingly, $G'(z) = 1/(1 + z/v)$ and $G''(z) = -1/v(1 + z/v)^2$.

### C. Resolving Scaling Ambiguity

The scaling ambiguity of IVA can be resolved directly by the commonly used minimal distortion principal (MDP) [32] as

$$\mathbf{W}^k \leftarrow diag\left(\left(\mathbf{W}^k\right)^{-1}\right) \mathbf{W}^k. \quad (16)$$

where diag(**X**) denotes the diagonal matrix with the same diagonal elements as those of matrix **X**. However, this strategy cannot be applied to FastIVE since the demixing matrix is unknown. Here we propose a row normalization approach.

Under the ideal condition that $\mathbf{W}^k = (\mathbf{H}^k)^{-1}$, (16) can be transformed into

$$\mathbf{W}^k \leftarrow diag(\mathbf{H}^k) \mathbf{W}^k. \quad (17)$$

Denote $\mathbf{h}_n^k$ as the $n$-th column vector of matrix $\mathbf{H}^k$ and $h_{ij}^k$ as the $i$-th row and $j$-th column element of $\mathbf{H}^k$, we can then write (17) in vector form as

$$\mathbf{w}_n^k \leftarrow h_{nn}^k \cdot \mathbf{w}_n^k, \quad (18)$$

Under the assumption of source independency, [25] shows that the mixing vector $\mathbf{h}_n^k$ can be calculated as

$$\mathbf{h}_n^k = \frac{\hat{\mathbf{C}}_x^k \mathbf{w}_n^k}{(\mathbf{w}_n^k)^H \hat{\mathbf{C}}_x^k \mathbf{w}_n^k}, \quad (19)$$

where $\hat{\mathbf{C}}_x^k$ is the sample-based estimate of $\mathbf{C}_x^k = E\left[\mathbf{x}^k (\mathbf{x}^k)^H\right]$. We can employ (18) to scale the demixing vector $\mathbf{w}_1^k$ after convergence.

### III. SIMULATIONS

#### A. Configurations

In the simulations, we use image model [33] to generate the mixtures, the room size is set to be 7 m × 5 m × 2.75 m and the reverberation time is 200 ms. As shown in Fig. 1, there are 6 possible source locations, and the center of an array with 6 microphones is located at [4, 1, 1.5] (m), with 1.25 cm interval between adjacent microphones. All speech signals (10 s long) are obtained from TIMIT database with 16 kHz sampling rate. The STFT is performed using a 2048-tap Hanning window with 512-tap shift. The proposed BSE with different signal models are named as FastIVE-SSL (spherical symmetric Laplacian model), FastIVE-GG (multivariate generalized Gaussian), and FastIVE-t (multivariate student's t model with freedom parameter $v$ set to be 4), respectively, and their performance is compared with the FIVE [26] and OGIVE-w [24] algorithms. The ILRMA with the desired speech manually extracted after separation is also included in simulations as a benchmark. The SIR improvement (SIRimp) [34] is utilized to evaluate the performance of different algorithms. Test samples are available at https://github.com/LiaoLele/audio-sample.

#### B. Evaluation results

We firstly evaluate the performance of different algorithms under different input SIRs with 2 speakers (sources 1 and 2) and 2 microphones (microphones 1 and 2). The SOI is source 1, and we change the input SIR of channel 1 by modulating the power of SOI. We obtain 300 pairs of mixtures at each input SIR and the SIRimp is calculated by averaging. We initialize $\mathbf{w}^k$ for all frequency bins with $(1\ 0)^T$ for all the FastIVE algorithms as well as FIVE, as described in Sec. II, and send the direction of arrival (DOA) of the SOI into OGIVE-w. Figure 2(a) depicts the performance of different algorithms when the desired speech is successfully extracted, and it can be seen that the ILRMA achieves stable SIRimp as expected. The FastIVE-t has comparatively better performance than the other two FastIVE algorithms, and overall the FastIVE algorithms achieve significantly better SIRimp than both the OGIVE-w and the FIVE since the interference signal model considerably deviates from the Gaussian distribution assumption. The extraction is determined as successful if the SIRimp is greater than 0 dB [24], and the success rate of different algorithms is shown in Fig. 2(b). It can be seen that the FastIVE can guarantee over 90% success rate when the input SIR is above 5 dB, significantly higher than the FIVE and OGIVE-w algorithms. Note that the success rate of the OGIVE-w is in accordance with the performance shown in [24], and the low success rate of the FIVE can be attributed to the fact that it cannot decide which source is of interest in the 2-source case.

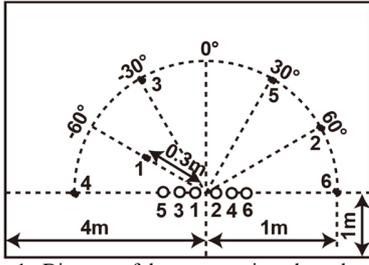

Fig. 1. Diagram of the source-microphone layout.

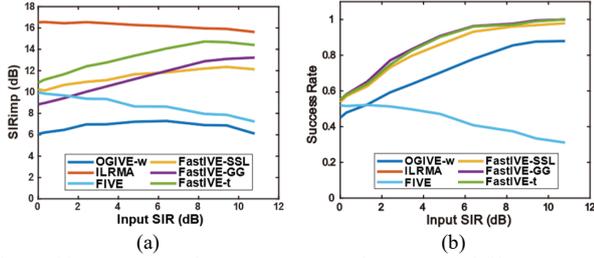

Fig. 2. (a) Average SIR improvements over input SIR and (b) Success rate over input SIR

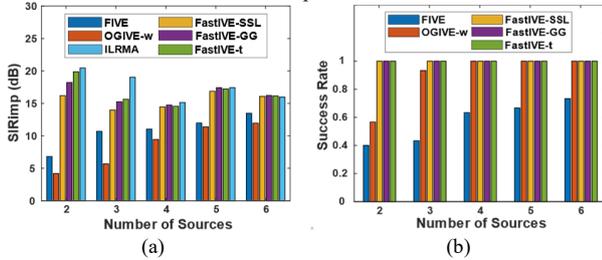

Fig. 3. (a) Average SIR improvements and (b) Success rate in fixed 2 channels case.

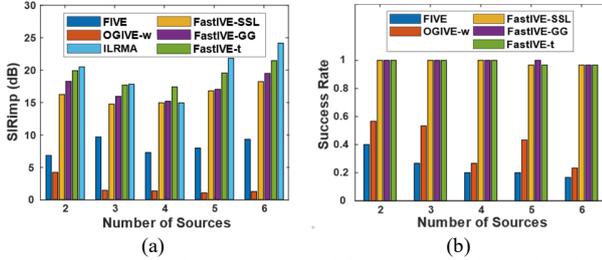

Fig. 4. (a) Average SIR improvements and (b) Success rate in varying channel number case.

To further validate the efficacy of the FastIVE algorithms, we evaluate the performance of different algorithms with varied number of sources and microphones, and the signal power of the SOI (source 1) is about 10 dB higher than each interfering speaker. In the first scenario, we fix the microphone number $M$ to be 2 (microphone 1 and 2 in Fig. 1) with source number $N$ varying from 2 to 6 (accordingly source 1-$N$ in Fig. 1 are chosen). We exploit all the 6 algorithms to extract the desired speech from 30 different mixtures in each case. Fig. 3(a) depicts the performance of different algorithms when the desired speech is successfully extracted, and it can be seen that overall the FastIVE algorithms perform better than the FIVE and the OGIVE-w methods, with SIRimp closer to that of the ILRMA. Note that both the FIVE and the OGIVE-w perform better with the increase of the source number, since the interference from more sources tends to have a Gaussian distribution. From the success rate shown in Fig. 3(b), it can be seen that all the FastIVE algorithms guarantee correct extraction in all source number cases. In the second scenario, the microphone number $M$ is set the same as the source number $N$. The average SIRimp of correct extractions is shown in Fig. 4(a). Unlike the 2-microphone case, the SIR improvement of the FIVE and the OGIVE-w does not show any advantage with the increase of the number of sources. However, the FastIVE algorithms, especially the FastIVE-t, make better use of more microphones and achieve better performance. The success rate shown in Fig. 4(b), also validates significant better extraction performance of the FastIVE algorithms.

The average run time with different microphone numbers (on a Linux computer with a 3.70GHz Intel Core i7 CPU) of 30 tests is listed in Table I. The FIVE has the lowest computational burden. The OGIVE-w consumes significantly longer runtime due to its slower convergence speed. Obviously the computational burden of the ILRMA increases sharply with the microphone number, since it has to separate all the sources. The time consumed by all the FastIVE algorithms, though longer than the FIVE, is significantly shorter than the ILRMA and the OGIVE-w, and its increase with the microphone number is limited. In summary, the FastIVE algorithms, especially the FastIVE-t, is a competitive choice of BSE in terms of performance and efficiency.

TABLE I
RUNTIME (IN SECONDS) OF VARIOUS ALGORITHMS

| Mic. number | 2 | 3 | 4 | 5 | 6 |
|---|---|---|---|---|---|
| FIVE | 0.07 | 0.10 | 0.13 | 0.16 | 0.19 |
| OGIVE-w | 18.95 | 20.39 | 22.11 | 22.66 | 24.02 |
| FastIVE-SSL | 0.50 | 0.83 | 0.84 | 1.32 | 1.88 |
| FastIVE-GG | 0.53 | 0.79 | 0.97 | 1.09 | 1.43 |
| FastIVE-t | 2.92 | 3.18 | 3.82 | 3.76 | 4.00 |
| ILRMA | 3.56 | 9.39 | 14.34 | 17.99 | 25.38 |

## IV. CONCLUSION

This paper proposes an efficient BSE method, FastIVE, by modifying IVA using the negentropy-based cost function with several different source prior models, with mild assumption that the SOI is from the dominant speaker in the mixture. We also design a scaling operation using only the SOI demixing vector to solve the scaling permutation problem. Simulations verify that the proposed algorithms, especially the FastIVE-t, can effectively extract the dominant speech with very high success rate, and the increase of their computational burden is limited with the increase of the microphone number.